# Validated research assessment based on highly cited researchers

Alonso Rodríguez-Navarro[a,b]*, Ricardo Brito[b]

[a] *Departamento de Biotecnología-Biología Vegetal, Universidad Politécnica de Madrid, Avenida Puerta de Hierro 2, 28040, Madrid, Spain*

[b] *Departamento de Estructura de la Materia, Física Térmica y Electrónica and GISC, Universidad Complutense de Madrid, Plaza de las Ciencias 3, 28040, Madrid, Spain*

\* Corresponding author: e-mail address: alonso.rodriguez@upm.es

**Abstract**

Bibliometrics provides accurate, cheap and simple descriptions of research systems and should lay the foundations for research policy. However, disconnections between bibliometric knowledge and research policy frequently misguide the research policy in many countries. A way of correcting these disconnections might come from the use of simple indicators of research performance. One such simple indicator is the number of highly cited researchers, which can be used under the assumption that a research system that produces and employs many highly cited researchers will be more successful than others with fewer of them. Here, we validate the use of the number of highly cited researchers (Ioannidis et al. 2020; PLoS Biol 18(10): e3000918) for research assessment at the country level and determine a country ranking of research success. We also demonstrate that the number of highly cited researchers reported by Clarivate Analytics is also an indicator of the research success of countries. The formal difference between the numbers of highly cited researchers according to Ionannidis et al. and Clarivate Analytics is that evaluations based on these two lists of highly cited researchers are approximately equivalent to evaluations based on the top 5% and 0.05% of highly cited papers, respectively. Moreover, the Clarivate Analytics indicator is flawed in some countries.

**1. Introduction**

A country's or institution's research system is a productive system that research policymakers need to understand in depth; bibliometrics provides an accurate, cheap and simple way of obtaining a scientific description of this system. Therefore, one of the most important applications of bibliometrics, if not the most important, is to provide a foundation for research policy (Garfield and Welljams-Dorof 1992).

When pursuing an improvement of a research system, it should be expected that policymakers will make use of current bibliometric knowledge and take decisions based thereon. However, overall, disconnections between current scientific knowledge and policy frequently cause misguided policy decisions (National Research Council 2012). This was detected in research policy in the EU 15 years ago (Dosi et al. 2006), and this disconnection is still present (Rodríguez-Navarro and Brito 2020a). Although the reasons for this may be complex, it appears that the best way to correct this disconnection may be to use a simple indicator that is easier for policymakers to interpret than metrics based on percentiles, probabilities, heavy tails or complex mathematical calculations.

A similar problem has been described previously as a tension between professional and citizen bibliometrics (Leydesdorff et al. 2016, p. 2129):

> the tension between simple but invalid indicators that are widely used (e.g., the h-index) and more sophisticated indicators that are not used or cannot be used in evaluation practices because they are not transparent for users, cannot be calculated, or are difficult to interpret.

Metrics based on Nobel Prizes and prestigious medals and awards (Charlton 2007) are simple and might be a solution, except for their scarcity.

**2. Highly cited researchers**

Instead of metrics based on a small number of prizes and awards, a metric based on the number of highly cited researchers (Docampo and Cram 2019) appears to be more convenient because their number can be high. The rational for this approach is that a research system that produces and employs many highly cited researchers (HCR) will probably be more successful than others with fewer of them (Bornmann and Bauer 2015a). This is similar to counting prizes and awards but at a much lower strict level, while retaining the notion that scientific progress is infrequent and that the success distribution of research results is heavy-tailed (Press 2013).

Under this notion, the number of HCR recorded by Clarivate Analytics from the Web of Science (WoS-HCR) has been used as an indicator of the research excellence of institutions and countries (Bauwens et al. 2011; Bornmann and Bauer 2015b, 2015a; Bornmann et al. 2018; Li 2016). However, the number of HCR can be counted in many ways, and an indicator of this type, as with any other metric, must be validated (Harnad 2009). The number of WoS-HCR has not yet been compared with validated metrics or peer review. Besides, it has been demonstrated that it is not a good predictor of the number of Nobel achievements (Rodríguez-Navarro 2011), and it has also been considered "a popular, albeit flawed, indicator of outstanding individual researchers" (Docampo and Cram 2019, p. 1011).

In 2019, Ioannidis et al. created another list of 105,026 HCR (IBB-HCR) drawn from 6,880,389 scientists who had published at least five papers during their career (Ioannidis et al. 2019). In 2020, this list was updated to include a total of 159,684 HCR, who lie within the top 2% of their main subfield discipline (176) according to the Science-Metrix classification (Ioannidis et al. 2020). Ioannidis et al. identified their HCR using an elaborate approach that resulted in a composite indicator specifically focused on capturing research success based on the number of citations together with other details of publications during the career of a researcher (Ioannidis et al. 2016). As explained above, assuming that the IBB-HCR were researchers with the most successful careers, the simple notion that this list can convey to policymakers is that the distribution of HCR across countries reflects the research success of each country and its potential to make future discoveries.

## 3. Aims and rationale of this study

The aim of the first part of this study was to provide a simple method for the research assessment of countries based on the number of IBB-HCR. However, for this purpose, the first step was to validate the use of this indicator so that the results of the assessment provided by the HCR metric were similar to those obtained from a peer assessment. Ioannidis et al. (Ioannidis et al. 2016) validated their composite indicator at the individual researcher level, but to create a country ranking based on the distribution of IBB-HCR, the method must be validated at this level.

To perform this validation, we used the data in the Leiden Ranking, which ranks universities according to the share of papers in four top percentiles (50, 10, 5, and 1) when the world's papers are ordered based on the number of citations (the US National Science Board also uses these indicators). Because in the evaluation of universities the number of papers in top percentiles was validated against the highest scores given in peer review (Rodríguez-Navarro and Brito 2020b; Traag and Waltman 2019), validation can be achieved by demonstrating that the number of IBB-HCR and the number of papers in a certain top percentile ($P_{\text{top } x\%}$) in the Leiden Ranking are highly correlated and take similar values across countries.

The second part of this study was aimed to investigate the relationship between the numbers of IBB- and WoS-HCR, because, as mentioned above, to be among the latter has been taken as an indicator of research excellence. A reasonable hypothesis was that the numbers of WoS- and IBB-HCR measure the same property of the research of a country (let us call it success or excellence) but at different levels of stringency because their global numbers are very different: 3,000–6,000 and around 160,000, respectively.

The obvious test of this hypothesis would be the same that is applied to validate of the number of IBB-HCR, that is, to find a certain top percentile ($P_{\text{top } y\%}$) in the Leiden Ranking for which the number of papers is similar to the number of WoS-HCR across countries. However, the global number of WoS-HCR is very small, and the hypothetical corresponding top percentile in the Leiden Ranking would be lower than the lowest top percentile (1%) reported by the Leiden Ranking.

In these circumstances, an alternative test for the hypothesis above was to demonstrate that, across countries, the numbers of WoS- and IBB-HCR are two points of the distribution function of success/excellence among researchers at two different levels of success/excellence, being about 50 times higher for the WoS-HCR than for the IBB-HCR.

Therefore, if the number of IBB-HCR is associated with a certain $P_{top\ x\%}$, as hypothesised above, it should be possible to associate the number of WoS-HCR with another $P_{top\ y\%}$. Consequently, because any $P_{top\ y\%}$ can be calculated from another $P_{top\ x\%}$ if the total number of papers is known (Rodríguez-Navarro and Brito 2021), our hypothesis predicts that the number of WoS-HCR can be calculated from the number of IBB-HCR; this is a hypothesis that can be tested.

**4. Data and methods**

As already explained above, our aim was to compare the number of IBB-HCR with the number of papers in a certain percentile across countries. However, neither Ioannidis et al. (2020) nor the Leiden Ranking provides data at the country level that could be directly compared and used for such a study. Therefore, to obtain comparable measures at country level, we identified the same universities in both the Leiden Ranking's and Ioannidis at al.'s lists and aggregated the data to obtain the country level.

For the IBB-HCR data, we downloaded Table S6 from the paper by Ioannidis et al. (Ioannidis et al. 2020); this is an Excel file containing the names and affiliations of 159,684 HCR who are ranked using a composite score that assesses scientists based on their career-long citation impact until the end of 2019. We also downloaded the Leiden Ranking 2020 (https://www.leidenranking.com/; August 21, 2020), which is also an Excel file, containing bibliometric data on 1,176 universities in six research fields and ten 4-year periods. To compare the Leiden data with the number of HCR according to Ioannidis et al., we selected the "All sciences" field and "fractional counting". For our study, the relevant data from the Leiden Ranking were the number of papers in the four top percentiles: 1, 5, 10 and 50 ($P_{top\ 1\%}$, $P_{top\ 5\%}$, $P_{top\ 10\%}$ and $P_{top\ 50\%}$).

Then, each university in the Leiden list was identified in the Ioannidis et al.'s Table 6. Most universities are listed in the Leiden Ranking under an English name, but this is not the norm in the Ioannidis et al.'s Table S6, thus the identification of the universities in this list was carried out based on the name both in the language of the country and in English. Furthermore, in the list by Ioannidis et al., many departments are recorded separately from the university to which they belong; in these cases, the number of HCR was obtained for the university as a whole by adding the numbers of HCR given separately for the university and its departments. We also checked the names of the departments carefully, as these are sometimes given in English and sometimes in the language of the country. The number of HCR in university hospitals was not included in the number of HCR for the university.

The WoS-HCR data were downloaded from https://clarivate.com/webofsciencegroup/thanks/?org=65406 (July 15, 2021). We downloaded a folder with Excel files containing the data corresponding to 2001 and to each year from 2014 to 2020. The data for years 2014–2020 include the names and affiliations, including the country, of the HCR.

For the second part of our study, i.e. the calculation of the number of WoS-HCR from the number of IBB-HCR, we used the method that is applied to calculate $P_{top\ y\%}$ from $P_{top\ x\%}$ (Rodríguez-Navarro and Brito 2019, 2021). In the first part of the study, we identified the $P_{top\ x\%}$ that can substitute for the number of IBB-HCR. If we found a $P_{top\ y\%}$ that could substitute for the number of WoS-HCR, the following equations would apply:

$$P_{top\ x\%} = P \cdot e_p^{(2-\lg x)} \qquad (1)$$

$$P_{top\ y\%} = P_{top\ x\%} \cdot e_p^{(\lg x - \lg y)} \qquad (2)$$

where $e_p$ is a constant that can be substituted by its proxy, the $P_{top\ 10\%}/P$ ratio (Rodríguez-Navarro and Brito 2019, 2021); this ratio can be calculated from the Leiden Ranking data used in the first part of our study (Supplementary Data 1).

Therefore, our hypothesis predicts that, across countries:

$$\text{Number of WoS-HCR} = \text{Number of IBB-HCR} \cdot (P_{top\ 10\%}/P)^{(lgx-lgy)} \quad (3)$$

and a comparison of the empirical and calculated numbers of WoS-HCR will reveal whether the numbers of WoS- and IBB-HCR are measuring the same property of a country's research but at two different levels of stringency.

**5. Results**

5.1. Selection of the Leiden data and validation of the number of IBB-HCR

From the process of matching universities in the Leiden Ranking and in the IBB-HCR list, after eliminating ambiguities and universities with zero records, we matched 1,111 universities in the two lists, belonging to 55 countries (Supplementary Data 1).

As described in Section 3, the first objective of our study was to validate the number of IBB-HCR against data in the Leiden Ranking 2020, which reports the results for ten 4-year periods, from 2006–2009 to 2015–2018. Because the number of IBB-HCR is a single value for each country, we had to select a particular percentile and period in order to calculate the correlation. When selecting the percentile, because the IBB-HCR represent about the top 2% of the total number of researchers, it seemed reasonable to use the top 5 or 1 percentile of the Leiden Ranking, $P_{top\ 1\%}$ or $P_{top\ 5\%}$. For statistical reasons $P_{top\ 5\%}$ was the best choice because $P_{top\ 1\%}$ is too low for some universities. Furthermore, we eventually found that the number of IBB-HCR and the $P_{top\ 5\%}$ in the 2006–2009 Leiden period had similar values (Supplementary Data 1; also see below).

The selection of the period had to be carried out empirically because the lists of the IBB-HCR correspond to the whole career of researchers while the numbers of highly cited papers in the Leiden Ranking correspond to fixed periods of four years. The IBB-HCR list includes the year in which each researcher published her/his first paper, thus making it possible to determine which citation period was dominant in the IBB-HCR. Figure 1 shows that 50% of the HCR published their first paper before 1984. Thus, many of the highly cited publications of the HCR were published before the first Leiden period (2006–2009), meaning that this period may be the most suitable for the validation since earlier periods are not available.

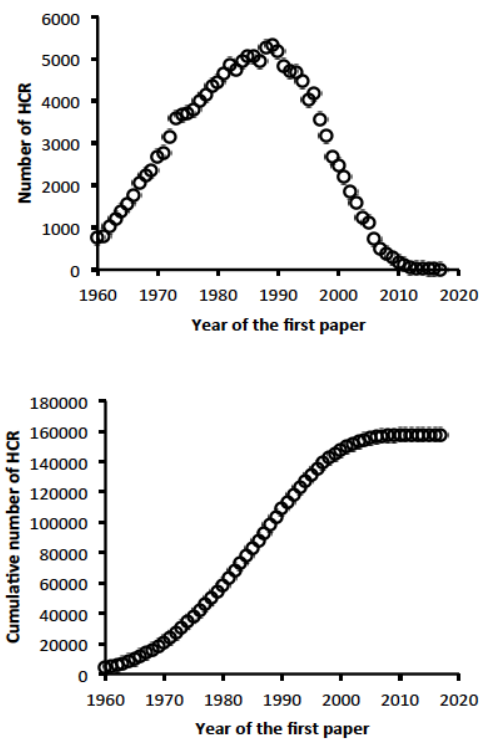

Figure 1. Distribution of papers in the ioannidis at al.'s list of HCR ordered by the year that is recorded for the first paper. The upper plot shows the number of HCR, while the lower plot shows the cumulative number of HCR

To make a correct selection of the period, we compared the number of IBB-HCR with the values of the $P_{top\ 5\%}$ for all periods (Supplementary Data 2). For the period 2006–2009, Spearman's rank correlation coefficient between the number of IBB-HCR and the $P_{top\ 5\%}$ was 0.97 (two-sided $p$-value, $6.8 \cdot 10^{-34}$); this coefficient decreased very little in subsequent periods. For the period 2006–2009, the Pearson correlation coefficient was 0.996 (two-sided $p$-value, $4.4 \cdot 10^{-58}$); excluding the USA because of its position as an outlier, the coefficient decreased to 0.97 (two-sided $p$-value $2.1 \cdot 10^{-33}$). In subsequent periods, excluding the USA, the Pearson coefficient decreased significantly to 0.72 (two-sided $p$-value, $1.0·10^{-9}$) in the period 2015–2018.

The scatter plot of countries constructed from the number of IBB-HCR and the $P_{top\ 5\%}$ was consistent with the tight correlation described. This conclusion was more evident in the scatter plot of ranks when counties were ordered from higher to lower numbers (Supplementary Data 3; Figure 2), although in a small number of countries (Qatar, Slovakia, Saudi Arabia, South Africa, and Turkey) the deviations were higher than the expected variability.

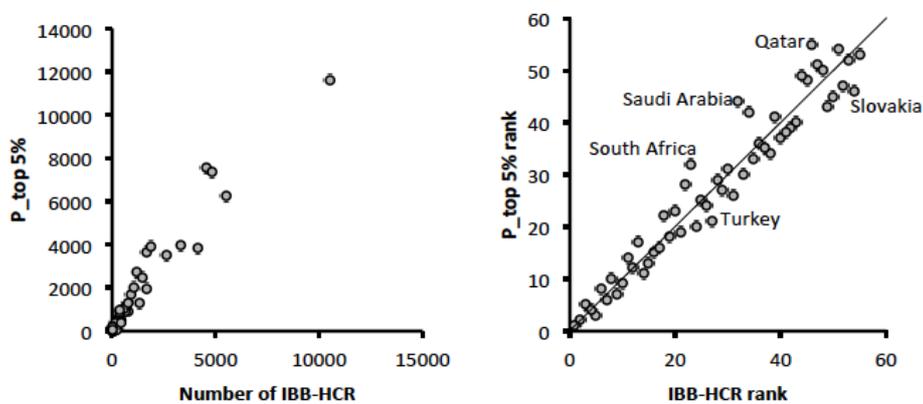

Figure 2. Scatter plot of countries: number of Ioannidis highly cited researchers (IBB-HCR) versus P_top 5%. Left panel: scatter plot of data, excluding the USA because its position as an outlier. Right panel: scatter plot of ranks, ordered from higher to lower values. The line with unity slope and zero intercept in right panel is drawn as a guide to the eye.

In summary, the high correlation coefficients between $P_{top\ 5\%}$ and the number of IBB-HCR in the 55 countries that we identified at university level in the lists of the Leiden Ranking and Ioannidis et al. strongly suggest that the number of IBB-HCR is a reliable indicator of research performance at the top 5% level of highly cited papers.

5.2. The IBB-HCR-based country ranking

Using a minimum threshold of 30 HCR, we listed 65 countries based on the numbers of IBB-HCR (Supplementary Data 4). The USA makes up 42.6% while the UK accounts for 9.4% of all IBB-HCR. Thus, these two countries account for more than 50% of all HCR, and the first 10 countries account for 80% of all IBB-HCR (Figure 3).

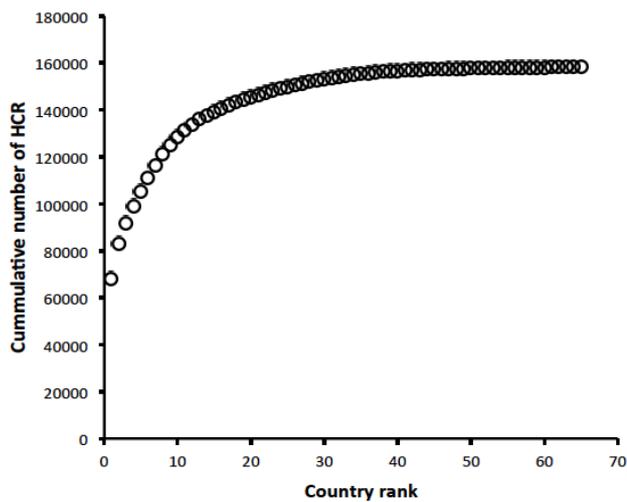

Figure 3. Cumulative number of HCR as a function of the ranking of the country (Supplementary Data 4). The first two countries, i.e. USA and UK, accounts for more than 50% of the total number of HCR

A country ranking based on the number of HCR is a size-dependent ranking that does not reflect size-independent characteristics such as research efficiency and commitment to research excellence. To reveal these country characteristics, the number of HCR can be normalised by the number of inhabitants or by the gross domestic product (GDP). Normalisation by the number of inhabitants shows that high research efficiency and commitment to research excellence are restricted to very few countries. The first three countries (Switzerland, Denmark and Sweden) have around 250 IBB-HCR per million inhabitants, but in the countries in positions 20, 21 and 22 (France, Italy and Greece) this figure decreases by a factor of four (Supplementary Data 4). Table 1 presents these data for 46 countries with more than 100 IBB-HCR.

A different ranking is obtained when normalising by GDP, but a similar decrease in the value of the indicator is observed. The top-ranked country in this case is UK with 5.3 IBB-HCR per billion US$ of GDP, while Spain and Portugal in positions 25 and 26 have only 1.6 HCR per billion US$ of GDP (Table 1).

The case of China merits particular attention. Although the number of IBB-HCR in China is currently at the same level as Australia or France, when normalised by the number of inhabitants or by the GDP, it ranks very low. However, it is worth noting that, between the first (2006–2009) and the last (2015–2018) periods of the Leiden

Ranking, the positions of China's universities have improved enormously (Supplementary Data 2).

Table 1. Number of Ioannidis et al's highly cited researchers: total and normalized by number of inhabitants and GDP

| Country | HCR | | Country | HCR per million inhabitants[a] | | Country | HCR per billion US$ GDP[a] |
|---|---|---|---|---|---|---|---|
| USA(USA) | 68016 | | CHE | 296.90 | | GBR | 5.30 |
| UK (GBR) | 15001 | | DKN | 257.12 | | SWE | 4.79 |
| Germany (DEU) | 8792 | | SWE | 247.69 | | DKN | 4.27 |
| Canada (CAN) | 7225 | | GBR | 224.44 | | CAN | 4.15 |
| Japan (JPN) | 6316 | | AUS | 214.50 | | ISR | 4.13 |
| Australia (AUS) | 5441 | | USA | 207.21 | | AUS | 3.90 |
| China (CHN) | 5272 | | NLD | 193.14 | | FIN | 3.86 |
| France (FRA) | 5048 | | CAN | 192.19 | | NZL | 3.84 |
| Italy (ITA) | 4008 | | FIN | 187.81 | | NLD | 3.69 |
| Netherlands (NLD) | 3350 | | ISR | 180.03 | | CHE | 3.48 |
| Sweden (SWE) | 2546 | | NOR | 177.08 | | USA | 3.17 |
| Switzerland (CHE) | 2546 | | NZL | 161.27 | | GRC | 3.16 |
| Spain (ESP) | 2290 | | SGP | 132.55 | | BEL | 2.65 |
| Israel (ISR) | 1630 | | BEL | 122.84 | | HKG | 2.42 |
| Denmark (DNK) | 1495 | | HKG | 116.95 | | NOR | 2.34 |
| India (IND) | 1491 | | IRL | 110.86 | | DEU | 2.28 |
| Belgium (BEL) | 1413 | | AUT | 108.33 | | AUT | 2.16 |
| South Korea (KOR) | 1350 | | DEU | 105.81 | | SVN | 2.10 |
| Taiwan | 1151 | | FRA | 75.28 | | SGP | 2.02 |
| Finland (FIN) | 1037 | | ITA | 66.47 | | ITA | 2.00 |
| Austria (AUT) | 962 | | GRC | 60.46 | | Taiwan | 1.88 |
| Norway (NOR) | 947 | | SVN | 54.59 | | FRA | 1.86 |
| Hong Kong (HKG) | 878 | | JPN | 50.02 | | HUN | 1.72 |
| New Zealand (NZL) | 803 | | Taiwan | 48.77 | | IRN | 1.68 |
| Singapore (SGP) | 756 | | ESP | 48.59 | | ESP | 1.64 |
| Poland (POL) | 726 | | PRT | 37.43 | | PRT | 1.61 |
| Russia (RUS) | 709 | | CZE | 31.39 | | ZAF | 1.53 |
| Greece (GRC) | 648 | | HUN | 28.86 | | IRL | 1.37 |
| Turkey (TUR) | 614 | | KOR | 26.11 | | CZE | 1.34 |
| Brazil (BRA) | 600 | | POL | 19.12 | | JPN | 1.25 |
| Ireland (IRL) | 547 | | ARE | 11.67 | | POL | 1.22 |
| South Africa (ZAF) | 536 | | SAU | 9.37 | | KOR | 0.82 |
| Iran (IRN) | 433 | | ZAF | 9.15 | | TUR | 0.81 |
| Portugal (PRT) | 385 | | TUR | 7.36 | | EGY | 0.72 |
| Czech Republic (CZE) | 335 | | CHL | 6.70 | | IND | 0.52 |
| Saudi Arabia (SAU) | 321 | | IRN | 5.22 | | CHL | 0.45 |
| Mexico (MEX) | 291 | | ROU | 5.16 | | MYS | 0.45 |
| Hungary (HUN) | 282 | | MYS | 5.10 | | RUS | 0.42 |
| Egypt (EGY) | 219 | | RUS | 4.91 | | SAU | 0.40 |
| Argentina (ARG) | 171 | | ARG | 3.81 | | ROU | 0.40 |
| Malaysia (MYS) | 163 | | CHN | 3.77 | | ARG | 0.38 |
| Thailand (THA) | 136 | | BRA | 2.84 | | CHN | 0.37 |
| Chile (CHL) | 127 | | MEX | 2.28 | | BRA | 0.32 |
| Slovenia (SVN) | 114 | | EGY | 2.18 | | ARE | 0.27 |
| United Arab Emirates (ARE) | 114 | | THA | 1.95 | | THA | 0.25 |
| Romania (ROU) | 100 | | IND | 1.09 | | MEX | 0.23 |

[a] World Bank data for 2019, except Taiwan

Relevant differences in research success between neighbouring advanced countries (Germany and France versus Switzerland and The Netherlands) have been revealed using several bibliometric approaches (Rodríguez-Navarro and Brito 2020a), and these results are reproduced by the number of IBB-HCR. The number of IBB-HCR per million inhabitants for Switzerland and the Netherlands is 297 and 193, while for Germany and France this figure is 106 and 75, respectively.

5.3. The numbers of WoS- and IBB-HCR are correlated

As mentioned above (Section 3), the second aim of our study was to demonstrate that the number of WoS-HCR could be calculated from the number of IBB-HCR across countries. For this purpose, our first step was to investigate whether these two numbers are correlated.

In the data downloaded from Clarivate, we selected the first year with sufficient information for our purposes (2014) because we validated the number of IBB-HCR by correlation with $P_{top\ 5\%}$ in the Leiden period of 2006–2009. In the downloaded list, the number of WoS-HCR was 3,216; in comparison, the number of IBB-HCR was 159,684. The large difference between these numbers implies that, as might be expected, countries with a small number of WoS-HCR had a highly variable number of IBB-HCR (Supplementary Data 5). Therefore, to enable a reliable comparison between the numbers of HCR in the two lists without the noise resulting from countries with small numbers of HCR, we deleted nine countries having one and eight countries having two WoS-HCR. We also eliminated China because it deviated too much from the general trend (a deviation explained by the above-mentioned rapid growth of research in China). These deletions reduced the number of countries from 49 to 33. Table 2 presents the numbers of WoS- and IBB-HCR for these 33 countries. The Spearman rank correlation coefficient between the numbers of WoS- and IBB-HCR was 0.80 (two-sided $p$-value, $2.1 \cdot 10^{-8}$), and the Pearson correlation coefficient when eliminating the USA data because of its position as an outlier was 0.97 (two-sided $p$-value, $9.3 \cdot 10^{-20}$).

Table 2. Number of highly cited researchers recorded by Clarivate Analytics (WoS-HCR) and by Ioannidis et al. (IBB-HCR)

| Country | WoS-HCR | IBB-HCR |
|---|---|---|
| Australia | 67 | 5441 |
| Austria | 17 | 962 |
| Belgium | 33 | 1413 |
| Brazil | 5 | 600 |
| Canada | 85 | 7225 |
| Denmark | 25 | 1495 |
| Finland | 13 | 1037 |
| France | 81 | 5048 |
| Germany | 157 | 8792 |
| Greece | 5 | 648 |
| Hong Kong | 18 | 878 |
| Iceland | 10 | 43 |
| India | 8 | 1491 |
| Iran | 11 | 433 |
| Ireland | 10 | 547 |
| Israel | 10 | 1630 |
| Italy | 47 | 4008 |
| Japan | 88 | 6316 |
| Netherlands | 72 | 3350 |
| New Zealand | 4 | 803 |
| Norway | 6 | 947 |
| Poland | 4 | 726 |
| Saudi Arabia | 28 | 321 |
| Singapore | 15 | 756 |
| South Africa | 8 | 536 |
| South Korea | 21 | 1350 |
| Spain | 40 | 2290 |
| Sweden | 27 | 2546 |
| Switzerland | 66 | 2546 |
| Taiwan | 12 | 1151 |
| Turkey | 10 | 614 |
| UK | 272 | 15001 |
| USA | 1612 | 68016 |

The scatter plot of the numbers of WoS- and IBB-HCR was consistent with the tight correlation described but also showed that several countries, especially those with small numbers of HCR, deviated from the general trend. This conclusion becomes more evident in the scatter plot of ranks when the counties are ordered from higher to lower numbers. Figure 4 shows that Saudi Arabia, Poland, India, Israel, Norway and New Zealand notably deviate from the line with unity slope and zero intercept. All these countries are situated in the upper-right part of the plot, which implies that they are situated in the lower part of the ranking.

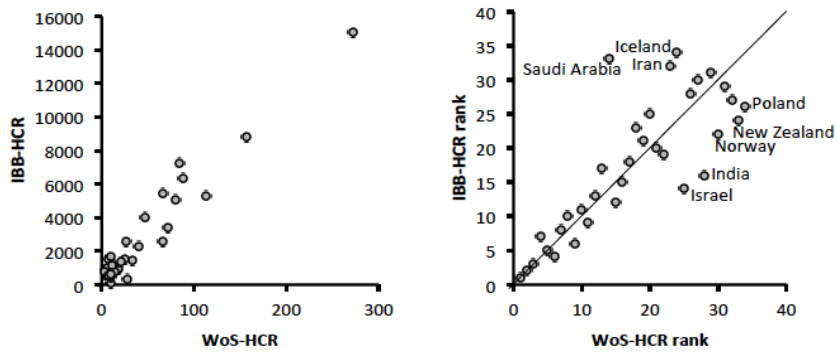

Figure 4. Scatter plot of countries: numbers of Clarivate Analytics (WoS-HCR) versus Ioannidis et al. highly cited researchers (IBB-HCR). Left panel: scatter plot of data, excluding the USA because of its position as an outlier. Right panel: scatter plot of ranks, ordered from higher to lower values. The line with unity slope and zero intercept in right panel is drawn as a guide to the eye.

5.4. Calculation of the number of WoS-HCR from the number of IBB-HCR

Next, we tested whether the number of WoS-HCR could be calculated from the number of IBB-HCR. For this calculation, we know that the number of IBB-HCR is associated with $P_{top\ 5\%}$ but not the $P_{top\ y\%}$ to which the WoS-HCR could be associated (Section 3). Considering the global numbers of WoS- and IBB-HCR of 3,216 and 159,684, respectively, if the total population of researchers from which these HCR were drawn were the same, the percentile corresponding to WoS-HCR would be 0.1.

However, the total populations of researchers from which the IBB- and WoS-HCR are drawn and the selection procedures cannot be compared easily. The IBB-HCR is restricted to researchers who have published at least five papers over their career up to 2019, while the population of researchers in the 2014 list of WoS-HCR are those who have published a paper in the period 2002–2012 (Docampo and Cram 2019). These considerations suggested that 0.1 was only a guiding percentile and we calculated several $P_{top\ y\%}$ from $P_{top\ 5\%}$ and compared the results with numbers of WoS-HCR. The results of these comparisons suggested that the best percentile to be associated to WoS-HCR was 0.05.

Next, according to this finding and based on Eq. 3, the calculated number of WoS-HCR in each country is equal to the number of IBB-HCR multiplied by the square (lg 5 – lg 0.05) of the $P_{top\ 10\%}/P$ ratio; this ratio for the university system could be obtained from the Leiden data (data aggregated at country level in Supplementary Data 1). The ratio for the university system might not be the ratio for the whole country's research system. However, conjecturing that the efficiency of research in the universities of a country cannot be very different from the efficiency of the whole research system of the country, we performed our calculations using the $P_{top10\%}/P$ ratio of the university system as the ratio for the research of the whole country. Table 3 presents the calculated values of WoS-HCR and the values of WoS-HCR reported by Clarivate Analytics for 31 countries (with respect to Table 2, Hong Kong (China in the Leiden Ranking) and Iceland are omitted because they are not recorded in Supplementary Data 1 and the $P_{top\ 10\%}/P$ ratios are unknown).

Table 3. Calculated and reported numbers of Clarivate Analytics highly cited researchers (WoS-HCR)

| Country | IBB-HCR | Leiden Ranking | | | WoS-HCR | |
|---|---|---|---|---|---|---|
| | | P | $P_{top\ 10\%}$ | $P_{top\ 10\%}/P$ | Calculated | Reported |
| Australia | 5441 | 71777 | 7673 | 0.107 | 62 | 67 |
| Austria | 962 | 17132 | 1796 | 0.105 | 11 | 17 |
| Belgium | 1413 | 29084 | 3324 | 0.114 | 18 | 33 |
| Brazil | 600 | 45905 | 2162 | 0.047 | 1 | 5 |
| Canada | 7225 | 112874 | 12497 | 0.111 | 89 | 85 |
| Denmark | 1495 | 18075 | 2273 | 0.126 | 24 | 25 |
| Finland | 1037 | 19636 | 1860 | 0.095 | 9 | 13 |
| France | 5048 | 65350 | 7340 | 0.112 | 64 | 81 |
| Germany | 8792 | 131726 | 14860 | 0.113 | 112 | 157 |
| Greece | 648 | 18945 | 1440 | 0.076 | 4 | 5 |
| India | 1491 | 33628 | 2303 | 0.068 | 7 | 8 |
| Iran | 433 | 22885 | 1314 | 0.057 | 1 | 11 |
| Ireland | 547 | 9961 | 1051 | 0.105 | 6 | 10 |
| Israel | 1630 | 27970 | 2583 | 0.092 | 14 | 10 |
| Italy | 4008 | 89071 | 7492 | 0.084 | 28 | 47 |
| Japan | 6316 | 133974 | 8580 | 0.064 | 26 | 88 |
| Netherlands | 3350 | 58205 | 7743 | 0.133 | 59 | 72 |
| New Zealand | 803 | 10330 | 1005 | 0.097 | 8 | 4 |
| Norway | 947 | 9990 | 1014 | 0.102 | 10 | 6 |
| Poland | 726 | 23259 | 889 | 0.038 | 1 | 4 |
| Saudi Arabia | 321 | 2575 | 130 | 0.051 | 1 | 28 |
| Singapore | 756 | 14957 | 1620 | 0.108 | 9 | 15 |
| South Africa | 536 | 10302 | 737 | 0.072 | 3 | 8 |
| South Korea | 1350 | 75930 | 4532 | 0.060 | 5 | 21 |
| Spain | 2290 | 67972 | 5854 | 0.086 | 17 | 40 |
| Sweden | 2546 | 38133 | 4043 | 0.106 | 29 | 27 |
| Switzerland | 2546 | 31237 | 4729 | 0.151 | 58 | 66 |
| Taiwan | 1151 | 42828 | 2938 | 0.069 | 5 | 12 |
| Turkey | 614 | 37903 | 2066 | 0.055 | 2 | 10 |
| UK | 15001 | 166306 | 22338 | 0.134 | 271 | 272 |
| USA | 68016 | 731994 | 106285 | 0.145 | 1434 | 1604 |

Abbreviations: Clarivate Analytics highly cited researchers, WoS-HCR; Ioannidis et al. highly cited researchers, IBB-HCR. Calculations: number of WoS-HCR = number of IBB-HCR · $(P_{top\ 10\%}/P)^2$

When comparing the reported and calculated numbers of WoS-HCR, the Spearman rank correlation coefficient was 0.83 (two-sided *p*-value 7.6·10$^{-9}$) and the Pearson coefficient excluding the USA due its position as an outlier was 0.97 (2 sided p-value 5.1·10$^{-18}$).

Overall, the scatter plots of data and ranks (Figure 5; with ranks ordered from the highest to lowest number of HCR) show good agreement between the calculated and observed numbers of WoS-HCR. For our purpose, the scatter plot of ranks, ranking the countries from higher to lower number of researchers, is more informative than the scatter plot of data because it shows more details of the comparison in countries with the smallest numbers of HCR. The scatter plot of ranks shows that many countries deviate very little from the line with unity slope and zero intercept. However, at least eight countries (Japan, Sweden, Saudi Arabia, South Korea, Israel, Spain Iran, and new Zealand) show notable deviations from this line.

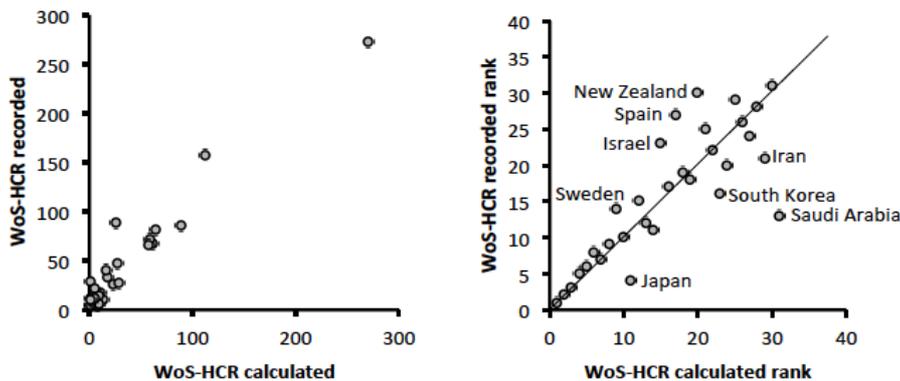

Figure 5. Plots of the number of highly cited researchers recorded by Clarivate Analytics (WoS-HCR) versus the homologous number calculated from the number of Ioannidis et al. highly cited researchers (see Table 3). Left panel: scatter plot of data, excluding the USA because of its position as an outlier. Right panel: scatter plot of ranks, ordered from higher to lower numbers. The line with unity slop and zero intercept in the right panel is drawn as a guide to the eye.

## 6. Discussion

Currently, research policy in many countries is disconnected from scientific/academic knowledge, in most cases because research policy uses indicators of research success that have been proved to be misleading for research assessment. The most widely studied case is the research policy of the EU, which is founded on a misleading European paradox. This paradox proposes excellent science and poor technology, based on an erroneous measurement of scientific excellence (Albarrán et al. 2010; Bonaccorsi 2007; Bonaccorsi et al. 2017; Dosi et al. 2006; Herranz and Ruiz-Castillo 2013; Rodríguez-Navarro and Narin 2018; Rodríguez-Navarro and Brito 2020a). Similar disconnections between policy and knowledge might occur in many countries around the world, misguiding the research policy.

The reasons for such disconnection between policy and academic knowledge are not clear, but the reason for the improper use of the impact factor for research assessment (Brito and Rodríguez-Navarro 2019), viz. its simplicity, might provide a hint: "*The JIF is wrong in so many ways, but it is so easy, a number that lets you rank scientists and their output in the same way as experimental data*" (Tregoning 2018, p. 345). In contrast to this simplicity, the skewness of science (Seglen 1992) and the heavy-tailed nature of the discovery distribution (Press 2013) mean that, until now, research assessment has been anything but simple for policymakers and, in general, for citizens who are not professional bibliometricians (Leydesdorff et al. 2016). It thus seems that if academic knowledge can offer policymakers a simple indicator of research success that is both reliable and statistically validated, its use in research policy might result attractive and help to connect policy with academic knowledge.

6.1. The number of IBB-HCR is a reliable indicator of research success

As mentioned above, the idea transmitted by the number of HCR is that their distribution across countries reflects the research success of each country and its potential to make future discoveries. According to this idea, the number of WoS-HCR has been used previously for research assessment (Bauwens et al. 2011; Bornmann and Bauer 2015b, 2015a; Bornmann et al. 2018; Li 2016).

In this study, we have demonstrated that the number of IBB-HCR is similar and shows a high correlation with the number of the top 5% of highly cited papers reported in the Leiden Ranking ($P_{top\ 5\%}$) for the period 2006–2009, in the field of "All sciences", and using "Fractional counting". The Pearson and Spearman rank correlation coefficients are very high, around 0.97, when eliminating the data for the USA data from the calculation of the Pearson coefficient because of its position as an outlier. Consistent with the notion that a large proportion of papers by the IBB-HCR were published before 2006 (Figure 1), our data also suggest that the correlation might have been even higher for periods before 2006–2009 (Supplementary Data 2), which are not reported in the Leiden Ranking. A Pearson correlation coefficient possibly higher than 0.97 for periods before 2006–2009 means an almost complete dependence of the two measures $P_{top\ 5\%}$ and number of HCR, which are obtained by two completely independent methods. They have in common only the fact that they measure the highest success, either in papers or researchers, respectively.

Previously, it has been shown that the numbers of papers in top percentiles correlate with the highest scores given in peer review in the Research Excellence Framework in the UK (Rodríguez-Navarro and Brito 2020b; Traag and Waltman 2019). Consequently, it can be assumed that $P_{top\ 5\%}$ is a validated measure of research success and that the number of IBB-HCR is another measure of it. It is worth noting that even if the number of papers in top percentiles had not been validated against peer review, the high correlation between IBB-HCR and $P_{top\ 5\%}$ would have also suggested that both parameters are measures of the same property of research.

In the scatter plot of ranks in Figure 2, countries as China, Saudi Arabia, Qatar, South Africa, and a few others deviate appreciably from the general trend. These countries are currently developing new research systems, and their rapid growth might make it difficult to evaluate them based on the number of IBB-HCR, who are selected considering a long research career. At least in the case of Saudi Arabia, the high number of HCR may be because of a policy of extensively hiring highly positioned foreign researchers (Bhattacharjee 2011).

The 65 studied countries with at least 30 IBB-HCR account for 99% of the IBB-HCR, but only 19 countries account for 90% of the IBB-HCR, indicating that a very low proportion of countries contribute significantly to global scientific progress (Figure 3).

However, some countries make a significant contribution because of their large size. Normalisation of the number of IBB-HCR by the population or GDP of the country shows that some of these countries that contribute significantly because of their size are not efficient: For example, Italy and Spain rank 9$^{th}$ and 14$^{th}$ by the number of IBB-HCR, respectively, but drop to 30$^{th}$ and 32$^{nd}$ position if their GDP is taken into consideration (Table 1). With respect to their GDP, these two countries are approximately four times less efficient that the leading countries.

The already mentioned differences in research success between countries that are economically and socially similar (the Netherlands, Switzerland, Germany, and France; Rodríguez-Navarro and Brito 2020a) are also shown by the number of IBB-HCR normalized by GDP. While the comparisons made in (Rodríguez-Navarro and Brito 2020a) have a complex bibliometric basis, the comparisons made on the basis of HCR have a simple basis and might be more convincing for policymakers and citizens without bibliometric knowledge.

The high correlations that we found when comparing the number of IBB-HCR with the Leiden data at country level—by aggregating universities—were not found at the university level (data not shown). For the universities of some countries, we found good correlations, whereas in other countries the correlations were low. We suspect that, at university level, correct matching of affiliations between the IBB-HCR and the Leiden data is difficult (Hottenrott et al. 2021; van Raan 2005), while affiliations at the country level can be matched without difficulties. However, in the research policy of countries, the assessment of the research of institutions based on the number of IBB-HCR should not face difficulties because, in most cases, policymakers should not have difficulties in identifying the correct affiliations of the IBB-HCR in their own country.

The IBB-HCR are classified into 22 scientific fields and 176 subfields (Ioannidis et al. 2020). Although we have not validated the number of HCR in these fields and subfields as indicators for research assessment, there is no reason to believe that they could not be used for this purpose. In this case and in the case of research assessment at the institutional level, the only reasonable limitation is that the number of HCR be sufficiently high that it is statistically reliable.

6.2. WoS-HCR is not always a reliable indicator of research success

As mentioned above, research success has been assessed using the number of WoS-HCR (Bauwens et al. 2011; Bornmann and Bauer 2015b, 2015a; Bornmann et al. 2018; Li 2016). We found that the numbers of IBB- and WoS-HCR (Table 2) are tightly correlated; this supports the notion that both numbers are similar measures of research success but at very different levels of stringency.

To further investigate this conclusion, under the rationale described above (Section 3), we associated the numbers of IBB- and WoS-HCR with $P_{top\ 5\%}$ and $P_{top\ 0.05\%}$, respectively, and calculated the number of WoS-HCR from the number of IBB-HCR as described above (Section 4). The calculated and reported numbers of WoS-HCR are similar and tightly correlated across countries (Figure 4). This result supports the hypothesis that the number of WoS-HCR can be calculated from the number of IBB-HCR, taking into account the research efficiency of each country as measured by the $P_{top\ 10\%}/P$ ratio (Table 3).

The scatter plot of ranks of the calculated and reported numbers of WoS-HCR (Figure 5) reveals a picture that is consistent with the tight correlation observed between the two numbers, although some countries show notable divergences. Excluding Japan and Sweden, the most important divergences occur in countries with small numbers of HCR. These divergences indicate that, in these countries, the probability that the number of WoS-HCR fails to measure research success correctly is higher than in countries with many HCR. Apart from this possibility, overall the deviations that occur in Japan, Sweden, South Korea, Israel, Spain and New Zealand seems to be a problem of the indicator. Because the number of IBB-HCR is validated (Section 6.1) and the recorded and calculated numbers of WoS-HCR are very similar in many countries, the conclusion is that the number of WoS-HCR fails to assess correctly the success of research in some specific countries.

Several reasons might explain this failure. Although study of these lies beyond the scope of this study, it should be noted that the WoS-HCR are selected from the papers that are highly cited using total counts (Docampo and Cram 2019) and many of these papers have a large number of authors, some of whom have not contributed to highly cited original research (Aksnes and Aagaar 2021). These peculiar highly cited authors might have a high weight in some countries.

In summary, the number of WoS-HCR may be a "flawed, indicator of outstanding individual researchers" ((Docampo and Cram 2019), p. 1011) but after aggregation, it is correct in many countries but wrong in some others as with individual researchers.

**7. Conclusions**

As outlined in the "Introduction" section, the formulation of simple indicators of research success might bring research policy and citizen bibliometrics closer to academic knowledge. A country's number of researchers in the HCR list of Ioannidis et al. (Ioannidis et al. 2020) is a simple indicator that policy makers can obtain by simply counting researchers in a list. Moreover, it transmits a simple idea of country research success: the more successful a country, the greater the number of successful researchers. In this study, we have validated the use of this indicator for research assessment at the country level and for all scientific fields together, but nothing raises the suspicion that it could not be used at the specific scientific field and institutional levels if the number of HCR is sufficiently high to be statistically reliable. The number of WoS-HCR is a reasonably indicator of research performance at a higher level of stringency than the number of IBB-HCR, but it is a flawed indicator in some countries.

**Funding**

This work was supported by the Spanish Ministerio de Economía y Competitividad '[grant number FIS2017-83709-R]'

**Supplementary data** can be requested to the authors

**References**

Aksnes, D W and Aagaar, K (2021), 'Lone geniuses or one among many? An exploratory study of contemporary highly cited researchers', *Journal of Data and Information Science,* 62, 41-66.

Albarrán, P, et al. (2010), 'A comparison of the scientific performance of the U.S. and the European Union at the turn of the 21st century', *Scientometrics,* 85, 329-44.

Bauwens, L, Mion, G, and Thisse, J-F (2011), 'The resistible decline of European science', *Recherches Économiques de Louvain,* 77, 5-31.

Bhattacharjee, Y (2011), 'Saudi universities offer cash in exchange for academic prestige', *Science,* 334, 1344-45.

Bonaccorsi, A (2007), 'Explaining poor performance of European science: institutions versus policies', *Science and Public Policy,* 34, 303-16.

Bonaccorsi, A, et al. (2017), 'Explaining the transatlantic gap in research excellence', *Scientometrics,* 110, 217-41.

Bornmann, L and Bauer, J (2015a), 'Which of the world's institutions employ the most highly cited resear An analysis of the data from highlycited.comchers?', *Journal of the Association for Information Science and Technology,* 66, 2146-48.

--- (2015b), 'Evaluation of the highly-cited researchers' database for a country: proposals for menaingful analyses on the example of Germany', *Scientometrics,* 105, 1997-2003.

Bornmann, L, Bauer, J, and Schlagberger, E M (2018), 'Highly cited researchers 2014 and 2015: An investigation of some of the world's most influencial scientific minds on the institutional and country level', *COLLNET Journal of Scientometrics and Information Management,* 12, 15-33.

Brito, R and Rodríguez-Navarro, A (2019), 'Evaluating research and researchers by the journal impact factor: Is it better than coin flipping?', *Journal of Informetrics,* 13, 314-24.

Charlton, B G (2007), 'Which are the best nations and institutions for revolutionay science 1987-2006? Analysis using a combined metric of Nobel prizes, Field medals, Lasker awards and Turing awards (NFLT metric)', *Medical Hypotheses,* 68, 1191-94.

Docampo, D and Cram, L (2019), 'Highly cited researchers: a moving target', *Scientometrics,* 118, 1011-25.

Dosi, G, Llerena, P, and Labini, M S (2006), 'The relationships between science, technologies and their industrial exploitation: An illustration through the myths and realities of the so-called 'European Paradox'', *Research Policy,* 35, 1450-64.

Garfield, E and Welljams-Dorof, A (1992), 'Citation data: their use as quantitative indicators for science and technology evaluations and policy-making', *Science and Public Policy,* 19, 321-27.

Harnad, S (2009), 'Open access scientometrics and the UK research assessment exercise', *Scientometrics,* 79, 147-56.

Herranz, N and Ruiz-Castillo, J (2013), 'The end of the "European Paradox"', *Scientometrics,* 95, 453-64.

Hottenrott, H, Rose, M E, and Lawson, C (2021), 'The rise of multiple institutional affiliations in academia', *Journal of the Association for Information Science and Technology,* 72, 1039-58.

Ioannidis, J P A, Boyack, K, and Wouters, P F (2016), 'Citation metrics: a primer on how (not) to normalize', *PLoS Biol,* 14(9), e1002542.

Ioannidis, J P A, Boyack, K W, and Baas, J (2020), 'Updated science-wide author databases of standardized indicators', *PLoS Biol,* 18(10), e000918.

Ioannidis, J P A, et al. (2019), 'A standardized citation metrics author database annotated for scientific field', *PLoS Biol,* 17(8), e3000384.

Leydesdorff, L, Wouters, P, and Bornmann, L (2016), 'Professional and citizen bibliometrics: complementarities and ambivalences in the development and use of indicators—a state-of-the-art report', *Scientometrics,* 109, 2129-50.

Li, J T (2016), 'What we learn from the shifts in highly cited data from 2001 to 2014?', *Scientometrics,* 108, 57-82.

National Research Council (2012), *Using Science as Evidence in Public Policy* (Washington, D.C.: The National Academies Press).

Press, W H (2013), 'What's so special about science (and how much should we spend on it?)', *Science,* 342, 817-22.

Rodríguez-Navarro, A (2011), 'Measuring research excellence. Number of Nobel Prize achievements versus conventional bibliometric indicators', *Journal of Documentation,* 67, 582-600.

Rodríguez-Navarro, A and Narin, F (2018), 'European paradox or delusion-Are European science and economy outdated?', *Science and Public Policy,* 45, 14-23.

Rodríguez-Navarro, A and Brito, R (2019), 'Probability and expected frequency of breakthroughs – basis and use of a robust method of research assessment', *Scientometrics,* 119, 213-35.

--- (2020a), 'Might Europe one day again be a global scientific powerhouse? Analysis of ERC publications suggests it will not be possiblewithout changes in research policy', *Quantitative Science Studies,* 1, 872-93.

--- (2020b), 'Like-for-like bibliometric substitutes for peer review: advantages and limits of indicators calculated from the ep index', *Research Evaluation,* 29, 215-30.

--- (2021), 'Total number of papers and in a single percentile fully describes reserach impact-Revisiting concepts and applications', *Quantitative Science Studies,* 2, 544-59.

Seglen, P O (1992), 'The skewness of sciance', *Journal of the American Society for information Science,* 43, 628-38.

Traag, V A and Waltman, L (2019), 'Systematic analysis of agreement between metrics and peer review in the UK REF', *Palgrave Communications,* 5, 29.

Tregoning, J (2018), 'How will you judge me if not by impact factor?', *Nature,* 558, 345.

van Raan, A F J (2005), 'Fatal attraction: Conceptual and methodological problems in the ranking of universities by bibliometric methods', *Scientometrics,* 62, 133-43.